&pdflatex

\documentclass[12pt]{iopart}
\usepackage{iopams}
\usepackage{setstack}
\usepackage{graphicx}

\newcommand{\cE}{\ensuremath{\mathcal{E}}}
\newcommand{\cP}{\ensuremath{\mathcal{P}}}
\newcommand{\cT}{\ensuremath{\mathcal{T}}}

\begin{document}

\rightline{preprint KCL-PH-TH-2013-19}

\title[Double-Scaling Limit of the O($N$)-Symmetric Anharmonic Oscillator]
{Double-Scaling Limit of the O($N$)-Symmetric Anharmonic Oscillator}

\author[Bender and Sarkar]{Carl~M~Bender$^{a,}$\footnote{{\footnotesize{\tt
email: cmb@wustl.edu}}} and Sarben Sarkar$^{b,}$\footnote{{\footnotesize{\tt
email: sarben.sarkar@kcl.ac.uk}}}}

\address{$^a$Department of Physics, Washington University, St. Louis MO 63130,
USA\\ $^b$Department of Physics, King's College London, Strand, London WC2R 2LS,
UK}

\date{today}

\begin{abstract}
In an earlier paper it was argued that the conventional double-scaling limit of
an O($N$)-symmetric quartic quantum field theory is inconsistent because the
critical coupling constant is negative and thus the integral representing the
partition function of the critical theory does not exist. In this earlier paper
it was shown that for an O($N$)-symmetric quantum field theory in
zero-dimensional spacetime one can avoid this difficulty if one replaces the
original quartic theory by its $\cP\cT$-symmetric analog. In the current paper
an O($N$)-symmetric quartic quantum field theory in one-dimensional spacetime
[that is, O($N$)-symmetric quantum mechanics] is studied using the Schr\"odinger
equation. It is shown that the global $\cP\cT$-symmetric formulation of this
differential equation provides a consistent way to perform the double-scaling
limit of the O($N$)-symmetric anharmonic oscillator. The physical nature of the
critical behavior is explained by studying the $\cP\cT$-symmetric quantum theory
and the corresponding and equivalent Hermitian quantum theory.
\end{abstract}


\submitto{\JPA}

\section{Introduction}
\label{s1}

Typically, the double-scaling limit of a quantum field theory is a correlated
limit characterized by a {\it universal} function of one parameter $\gamma$,
which is a combination of the original parameters in the Hamiltonian or
Lagrangian \cite{R1}. This universal function, which is {\it entire} (analytic
for all $\gamma$), reveals the essential features of the theory while being
insensitive to specific details. In quantum field theory the correlated limit of
an O($N$)-symmetric vector model represents a discretized branched polymer
\cite{R2}.

This paper examines the double-scaling limit of an O($N$)-symmetric quartic
quantum field theory in one-dimensional spacetime. As in our previous paper on
the double-scaling limit in zero-dimensional spacetime \cite{R3}, we argue here
that the conventional double-scaling limit in one-dimensional spacetime is
inconsistent because the critical value of the coupling constant is negative. We
then show that the double-scaling limit {\it is} consistent for the associated
$\cP\cT$-symmetric quantum field theory. However, we do not follow the
procedures used in Ref.~\cite{R3} because this would require the introduction
of a functional-integral representation for the partition function, which is an
infinite-dimensional integral. Analyzing Stokes wedges in this context is
complicated and difficult. Fortunately, the one-dimensional quantum field theory
is equivalent to an O($N$)-symmetric quantum-mechanical anharmonic oscillator,
and thus the theory is described by a Schr\"odinger equation \cite{R4}. It is
easier and more physically transparent to study the properties of a wave
function than the boundary conditions and convergence of an infinite-dimensional
integral.

The conventional coordinate-space anharmonic-oscillator Hamiltonian is
\begin{equation}
H=-\sum_{j=1}^{N+1}\frac{\partial^2}{\partial x_j^2}+\frac{\mu^2}{2}
\sum_{j=1}^{N+1}x_j^2+\frac{\lambda}{4}\left(\sum_{j=1}^{N+1}x_j^2\right)^2.
\label{e1}
\end{equation}
To obtain the double-scaling limit of this theory, we begin by constructing the
uncorrelated large-$N$ expansion in powers of $1/N$. This expansion is a
divergent asymptotic series in powers of $1/N$ and has the general form
\cite{R5}
\begin{eqnarray}
\sum_{k=0}^\infty a_kN^{-k},
\label{e2}
\end{eqnarray}
where the coefficients $a_k$ have a nontrivial dependence on the parameters
$\mu$ and $\lambda$ in the Hamiltonian. Only the first few terms in the $1/N$
expansion are calculable with reasonable effort.

To study the Hamiltonian (\ref{e1}) for $N>>1$, we rewrite the time-independent
Schr\"odinger equation $H\psi=E\psi$ in polar coordinates by substituting
$\sum_{j=1}^{N+1}x_j^2=Nr^2$ (see Ref.~\cite{R6}) and we let $\lambda=g/N$ and
$\cE\equiv E/N$. We then seek a spherically-symmetric solution $\psi(r)$, which
obeys the differential equation
\begin{equation}
-\frac{1}{N^2}\psi''(r)-\frac{1}{Nr}\psi'(r)+\frac{\mu^2}{2}r^2\psi(r)+\frac{g}
{4}r^4\psi(r)=\cE\psi(r).
\label{e3}
\end{equation}
We can convert (\ref{e3}) to the form of a radial Schr\"odinger equation by
making the change of variable $\Phi(r)=r^{N/2}\psi(r)$. The resulting
Schr\"odinger equation has the form
\begin{equation}
-\frac{1}{N^2}\Phi''(r)+V(r)\Phi(r)=\cE\Phi(r),
\label{e4}
\end{equation}
where for large $N$
\begin{equation}
V(r)=\frac{1}{4r^2}+\frac{\mu^2}{2}r^2+\frac{g}{4}r^4.
\label{e5}
\end{equation}
(Originally, the coefficient of $r^{-2}$ in the potential is $\frac{1}{4}-
\frac{1}{2N}$, so for large $N$ we replace this coefficient by $\frac{1}{4}$.)
Note that $1/N$ plays the role of $\hbar$ in the Schr\"odinger equation
(\ref{e4}) and thus the large-$N$ expansion (\ref{e2}) is just the standard
semiclassical (WKB) expansion.

To find the double-scaling limit of the WKB expansion we must take the limits
$N\to\infty$ and $g\to g_{\rm crit}$ in a {\it correlated} fashion. In this
limit the WKB series (\ref{e2}) undergoes a transmutation in which all terms
become of comparable size; the series is no longer dominated by early terms and
$N^{-k}$ in the $k$th term is balanced by a large coefficient $a_k$ \cite{R1}.
In this correlated limit the perturbation series still diverges. The question is
whether we can apply a summation procedure such as Borel summation to the
series. In Ref.~\cite{R3} we found that in the double-scaling limit the series
was {\it not} Borel summable because the terms did not alternate in sign, and
the lack of Borel summability was the indication that the conventional quartic
O($N$)-symmetric theory did not in fact have a double-scaling limit.

\begin{figure}[h!]
\begin{center}
\includegraphics[scale=0.32]{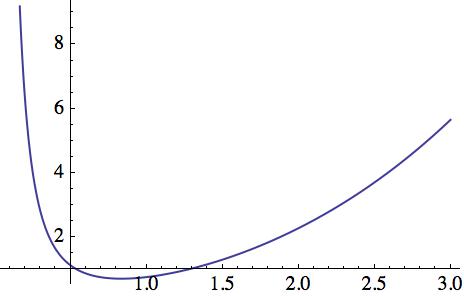}
\hspace{.5cm}
\includegraphics[scale=0.26]{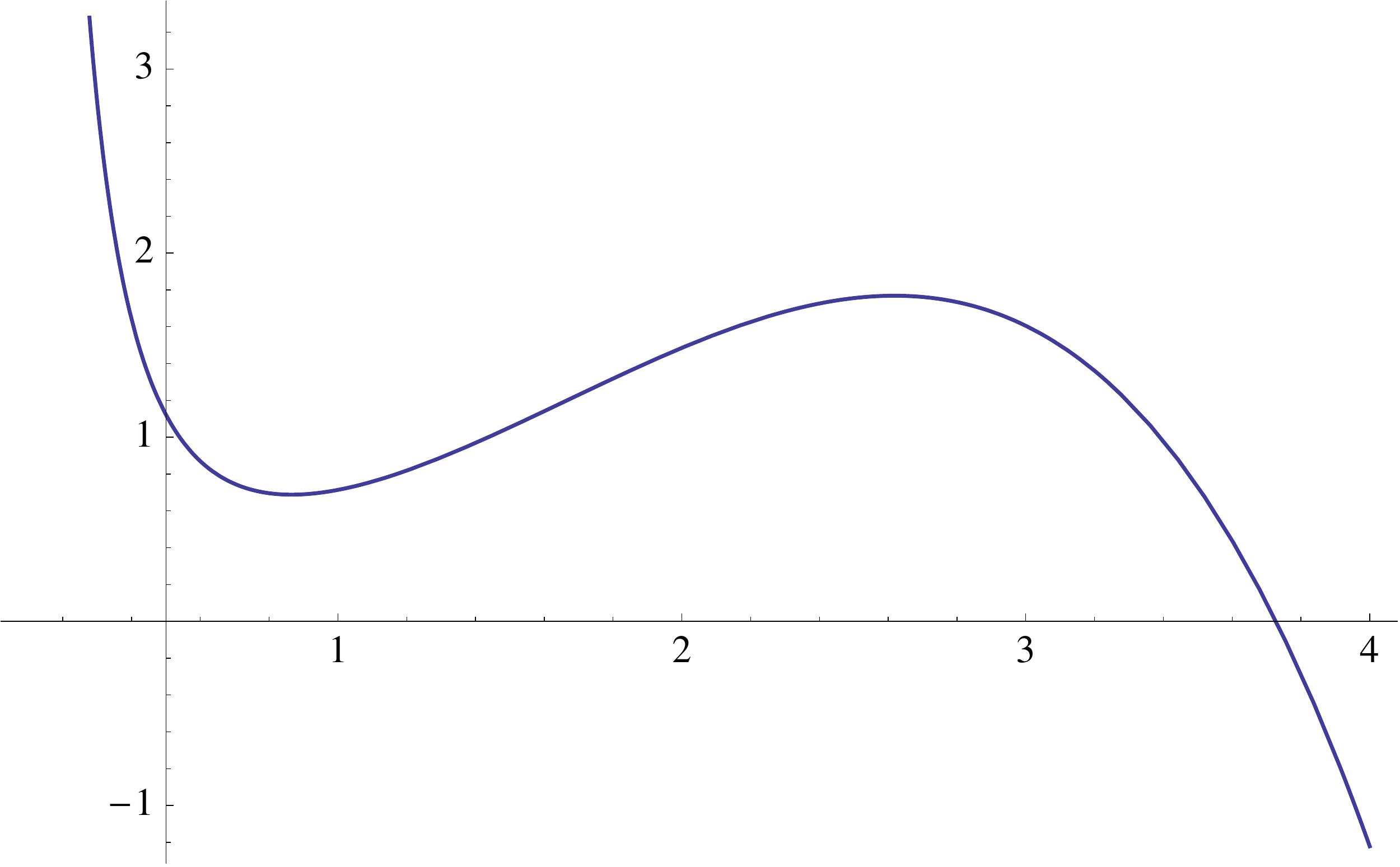}
\end{center}
\caption{Plot of the potential $V(r)$ in (\ref{e5}) for $\mu=1$ and for two
values of $g$. Left panel: $g=g_{\rm crit}+0.6$, where $g_{\rm crit}=-0.544331$;
the potential rises as $r\to\infty$ and thus it confines bound states. Right
panel: $g=g_{\rm crit}+0.4$; the potential falls as $r\to\infty$, so particles
tunnel out to $\infty$ and states become quasi-stable.}
\label{F12}
\end{figure}

\begin{figure}[h!]
\begin{center}
\includegraphics[scale=0.26]{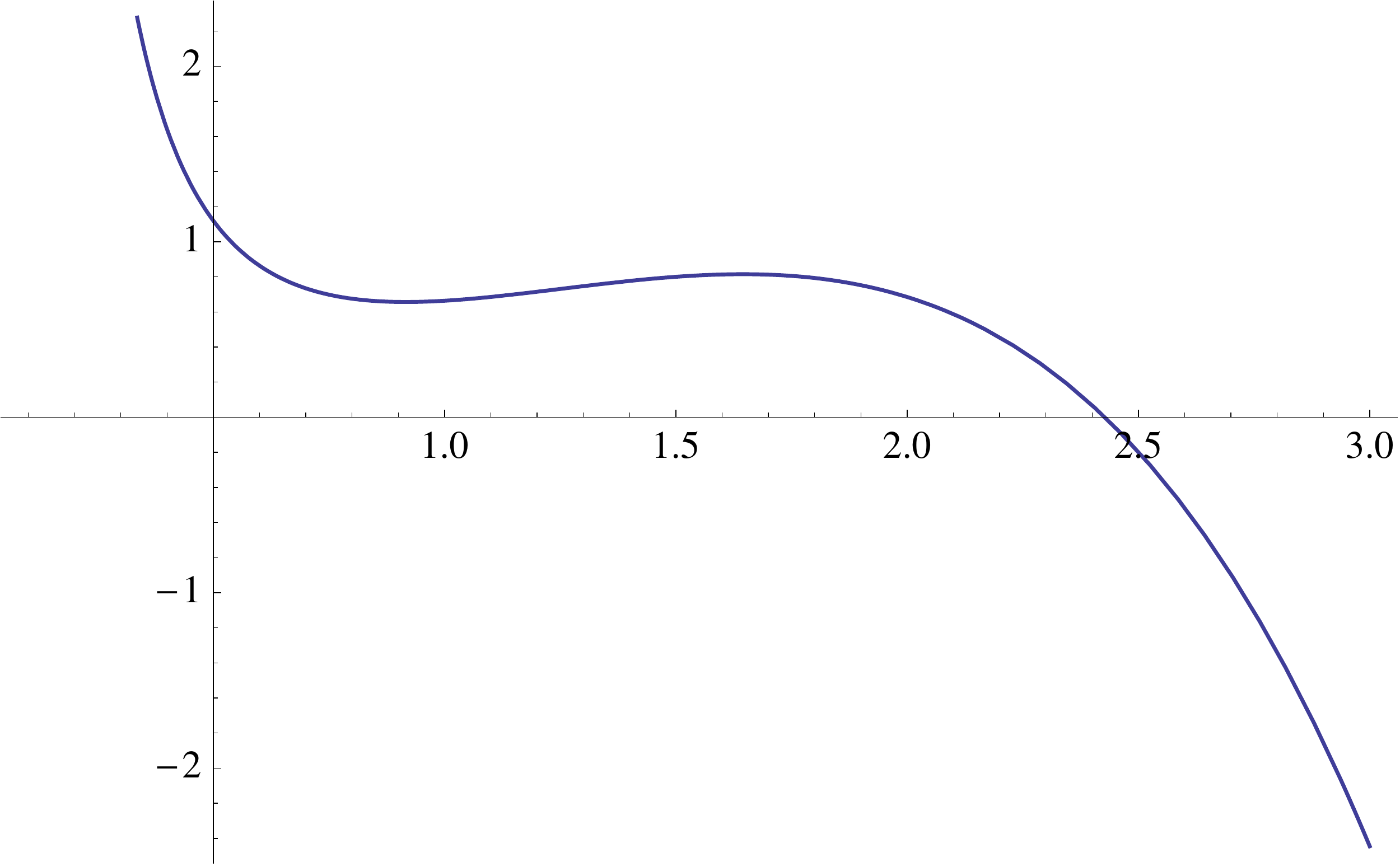}
\hspace{.5cm}
\includegraphics[scale=0.26]{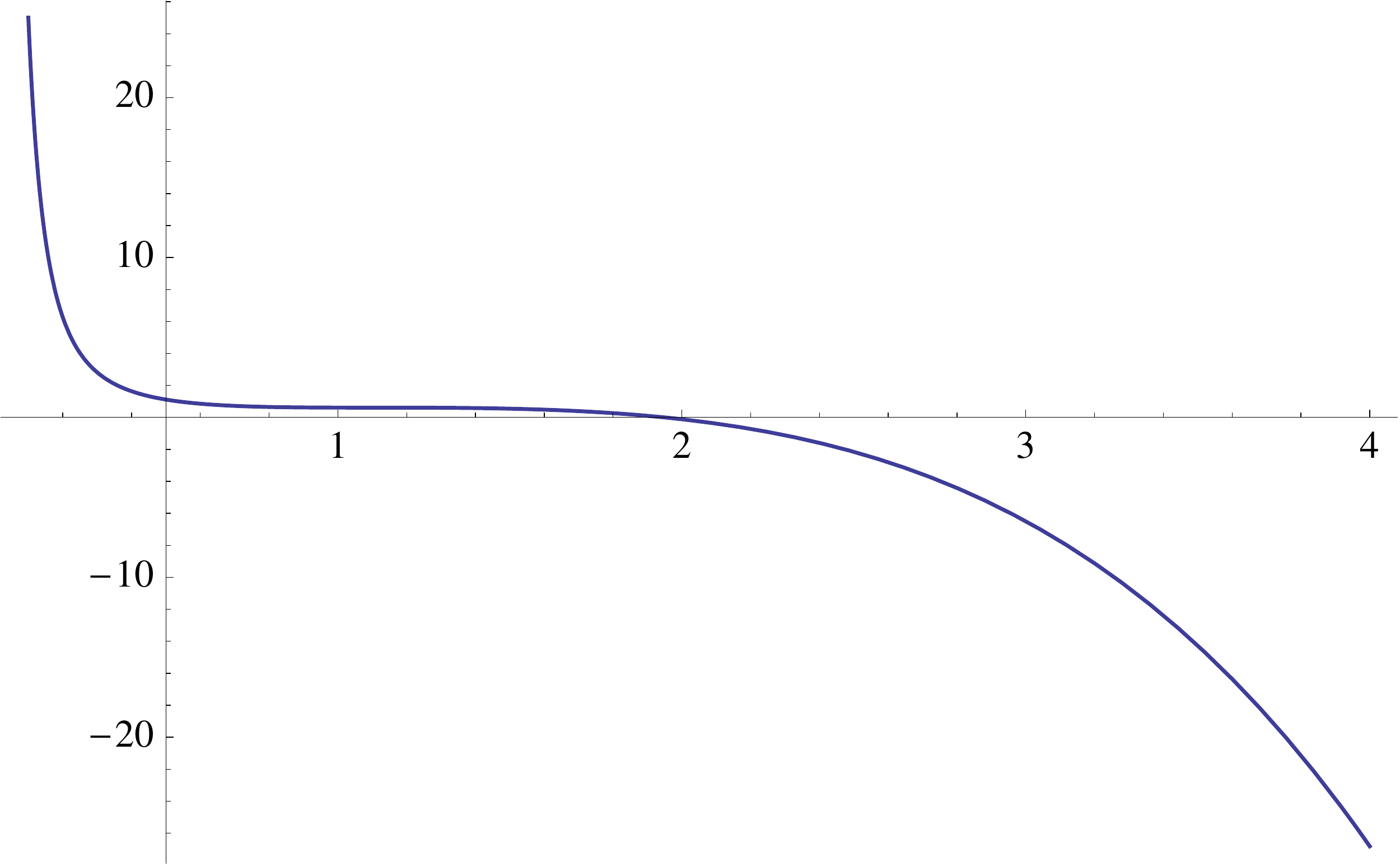}
\end{center}
\caption{The potential $V(r)$ in (\ref{e5}) for $\mu=1$ and for two values of
$g$. Left panel: $g=g_{\rm crit}+0.2$, where $g_{\rm crit}=-0.544331$; for this
value of $g$ the potential just barely has quasi-stable states. Right panel: $g=
g_{\rm crit}$; particles move unimpeded off to $r=\infty$ without tunneling.
Note that at the critical value of $g$ the maximum and minimum of the potential
coalesce into an inflection point.}
\label{F34}
\end{figure}

We emphasize that the solution to the Schr\"odinger equation (\ref{e4}) is
required to be normalizable and thus it must vanish as $r\to\infty$. However,
the correlated limit is defined by the pair of equations \cite{R7}
\begin{equation}
V'(r)=-\frac{2}{r^3}+\mu^2r+gr^3=0\quad{\rm and}\quad
V''(r)=-\frac{6}{r^4}+\mu^2+3gr^2=0.
\label{e6}
\end{equation}
Solving these equations simultaneously, we find that the critical value of $g$
is {\it negative}:
\begin{equation}
g_{\rm crit}=-(2/3)^{3/2}\mu^3\approx-0.544331\mu^3.
\label{e7}
\end{equation}
This result reveals the problem with taking the double-scaling limit of the
quantum theory whose potential $V(r)$ is given in (\ref{e5}). When $g$ is
positive, $V(r)$ confines bound states and the eigenfunction is normalizable
(Fig.~\ref{F12}, left panel). However, as $g$ moves downward toward its critical
value (\ref{e7}), the potential turns over as $r\to\infty$. When $g$ becomes
negative, the quartic term in $V(r)$ allows particles to tunnel out to $r=
\infty$. The states of the theory become quasi-stable (Fig.~\ref{F12}, right
panel) and the solutions to the Schr\"odinger equation (\ref{e4}) are not
normalizable. As $g$ continues to decrease towards its critical value, the
potential barrier decreases in size (Fig.~\ref{F34}, left panel) and at the
critical value the potential barrier {\it disappears entirely} (Fig.~\ref{F34},
right panel). Thus, there are no longer any quasi-stable states; quantum
particles flow unimpeded and without tunneling out to infinity. This gives a
physical picture of the transition that occurs at the double-scaling limit and
demonstrates the physical problem with taking the double-scaling limit of the
conventional Hermitian O($N$)-symmetric anharmonic oscillator.

We show in this paper that while the conventional O($N$)-symmetric anharmonic
oscillator does not possess a double-scaling limit, the corresponding $\cP
\cT$-symmetric O($N$)-symmetric anharmonic oscillator {\it does} have a
double-scaling limit. Unlike the conventional anharmonic oscillator, the $\cP
\cT$-symmetric anharmonic oscillator has a {\it negative} quartic term in the
potential. However, it still possesses normalizable eigenfunctions because the
Schr\"odinger equation for the $\cP\cT$-symmetric theory is posed in the complex
plane and is normalized in appropriate complex Stokes wedges. We show how to
construct the $\cP\cT$-symmetric anharmonic oscillator in Sec.~\ref{s2}. Then by
constructing the equivalent Hermitian quantum theory in Sec.~\ref{s3}, we
explain the physical transformation that occurs at the critical value of the
coupling constant in the double-scaling limit. Concluding remarks are given in
Sec.~\ref{s4}.

\section{$\cP\cT$-symmetric formulation of the O($N$)-symmetric anharmonic
oscillator}
\label{s2}

Following the procedure for the case of the O($N$)-symmetric zero-dimensional
quantum field theory in Ref.~\cite{R3}, we confront the problem of negative
$g_{\rm crit}$ [see (\ref{e7})] by considering the family of O($N$)-symmetric
$\cP\cT$-symmetric Hamiltonians
\begin{equation}
H=-\sum_{j=1}^{N+1}\frac{\partial^2}{\partial x_j^2}+\frac{1}{2}\sum_{j=1}^{N+1}
x_j^2+\frac{\lambda}{2+\epsilon}\left(\sum_{j=1}^{N+1}x_j^2\right)i^\epsilon
\left(\sum_{j=1}^{N+1}x_j^2\right)^{\epsilon/2},
\label{e8}
\end{equation}
where $\epsilon\geq0$ is a real parameter. At $\epsilon=0$ this Hamiltonian
describes an $(N+1)$-component harmonic-oscillator, but as $\epsilon$ approaches
$2$ the Hamiltonian becomes quartic and a negative quartic term arises
naturally. As in Sec.~\ref{s1} (and as in Ref.~\cite{R3}), we introduce polar
coordinates $\sum_{j=1}^{N+1}x_j^2=Nr^2$ and let $g=\lambda N^{\epsilon/2}$. The
spherically symmetric solution $\Phi(r)=r^{N/2}\psi(r)$ satisfies the
differential-equation eigenvalue problem
\begin{equation}
-\frac{1}{N^2}\Phi''(r)+V_\epsilon(r)\Phi(r)=\cE\Phi(r),
\label{e9}
\end{equation}
where $\cE$ is the eigenvalue and
\begin{equation}
V_\epsilon(r)\equiv\left(\frac{1}{4}-\frac{1}{2N}\right)\frac{1}{r^2}+\frac{1}
{2}r^2+\frac{g}{2+\epsilon}r^2(ir)^{\epsilon}.
\label{e10}
\end{equation}

Let us analyze the differential equation (\ref{e9}) Rlocally near the origin by
using Frobenius theory. In the vicinity of $r=0$ this differential equation is
approximated by the equidimensional equation
\begin{equation}
-\frac{1}{N^2}\Phi''(r)+\left(\frac{1}{4}-\frac{1}{2N}\right)\frac{1}{r^2}\Phi
(r)=0.
\label{e11}
\end{equation}
As $r\to0$, $\Phi(r)$ is approximated by $r^\alpha$, where $\alpha$ is the
Frobenius index. The possible values of $\alpha$ are $N/2$ and $1-N/2$. Thus,
there is one solution that is finite at the origin and another that is
divergent. We then make the assumption that $N$ is even so that $r^{N/2}$ is
single-valued at the origin. This allows us to extend the range of $r$ from the
positive-$r$ axis to the whole real-$x$ axis. The case of even $N$ is special
because the parity reflection operator in ($N+1$)-dimensional space, which has
the effect of changing the sign of all spatial components of the coordinate
vector, is distinct from a rotation. [When $N$ is odd, parity reflection in $(N+
1)$-dimensional space is just a rotation and thus there is no distinct parity
operator.] The existence of a distinct $\cP$ operator is crucial in a $\cP
\cT$-symmetric theory. (It should be noted that a similar result regarding the
dimension of space, namely, that $N$ is odd, was found in Ref.~\cite{R3} using
a different argument.)

We then let $\epsilon$ approach $2$ in order to obtain the quartic (anharmonic)
theory. In this limit the boundary conditions on the differential equation
(\ref{e9}) extended from $r$ space to $x$ space, which are imposed in Stokes
wedges that rotate downward off the real-$x$ axis and into the complex-$x$
plane, are treated as explained in Ref.~\cite{R8}. This procedure defines a $\cP
\cT$-symmetric quantum theory that has a positive real spectrum. In the next
section we demonstrate rigorously the positivity of the spectrum at $\epsilon=2$
by constructing an exact equivalent Hermitian quantum theory.

\section{Construction of an equivalent Hermitian theory}
\label{s3}

We will now prove that when the positive variable $r$ is replaced by the complex
variable $x$, the $\cP\cT$-symmetric Schr\"odinger equation (\ref{e9}) has a
positive real spectrum when $\epsilon=2$. In doing so we will understand in 
physical terms how to interpret the critical behavior in the double-scaling
limit.

We begin with the large-$N$ $\cP\cT$-symmetric Schr\"odinger equation
\begin{equation}
-\frac{1}{N^2}\psi''(x)+\left(\frac{1}{4x^2}
+\frac{1}{2}x^2-\frac{g}{4}x^4\right)\psi(x)=\cE\psi(x)
\label{e12}
\end{equation}
and we substitute $\psi(x)=x^\beta\phi(x)$. Hence, $\psi''(x)=\beta(\beta-1)x^{
\beta-2}\phi(x)+2\beta x^{\beta-1}\phi'(x)+x^\beta\phi''(x)$. Then, if we let
$\beta(\beta-1)=N^2/4$, we obtain
\begin{equation}
\left(-\frac{1}{N^2}\frac{d^2}{dx^2}-\frac{2\beta}{N^2x}\frac{d}{dx}
+\frac{1}{2}x^2-\frac{g}{4}x^4\right)\phi(x)=\cE\phi(x).
\label{e13}
\end{equation}

Next, generalizing the work of Ref.~\cite{R9} for transforming a quartic theory
with a negative coupling constant to one with a positive coupling constant, we
substitute $x=-2i\sqrt{1+it}$. Note that as $t$ ranges along the real-$t$ axis
from $t=-\infty$ to $t=\infty$, $x$ traces a path in the complex-$x$ plane that
originates and terminates in both Stokes wedges. This transformation gives
$$\frac{d}{dx}=\frac{ix}{2}\frac{d}{dt}\qquad{\rm and}\qquad
\frac{d^2}{dx^2}=\frac{i}{2}\frac{d}{dt}+(1+it)\frac{d^2}{dt^2},$$
and we obtain the following differential equation on the real-$t$ axis:
$$-\frac{1+it}{N^2}\phi''(t)-i\frac{1+2\beta}{2N^2}\phi'(t)-\left[
2(1+it)+4g(1+2it-t^2)\right]\phi(t)=\cE\phi(t).$$
We then take a Fourier transform of this equation according to $\tilde\phi(s)
\equiv\int_{-\infty}^\infty dt\,e^{its}\phi(t)$. The resulting equation for
$\tilde\phi$ is
$$\left(\frac{3-2\beta}{2N^2}s+\frac{1}{N^2}s^2+\frac{1}{N^2}s^2\frac{d}{ds}-2-2
\frac{d}{ds}-4g-8g\frac{d}{ds}-4g\frac{d^2}{ds^2}\right)\tilde\phi(s)=\cE
\tilde\phi(s).$$

\begin{figure}[h!]
\begin{center}
\includegraphics[scale=0.26]{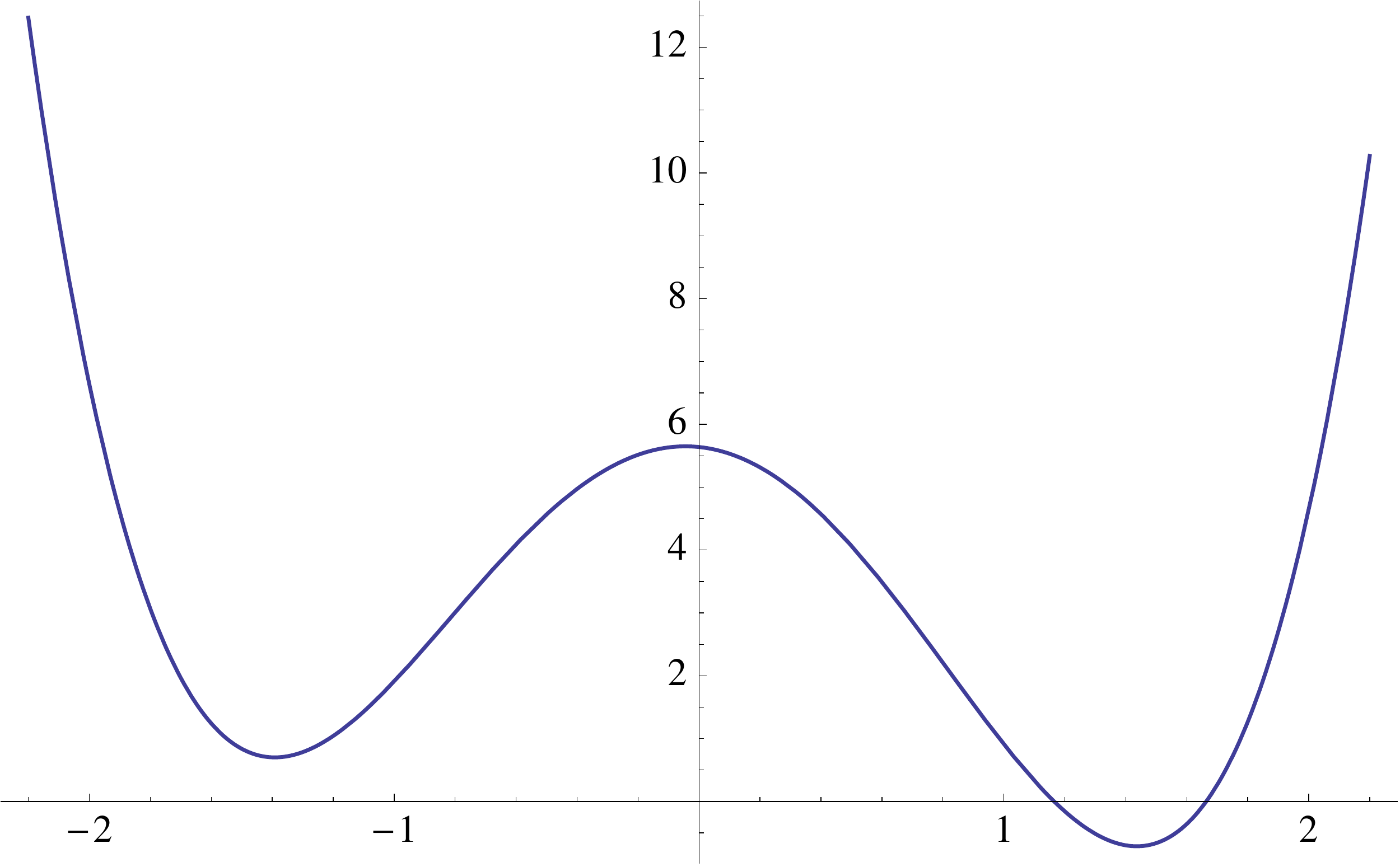}
\hspace{.5cm}
\includegraphics[scale=0.26]{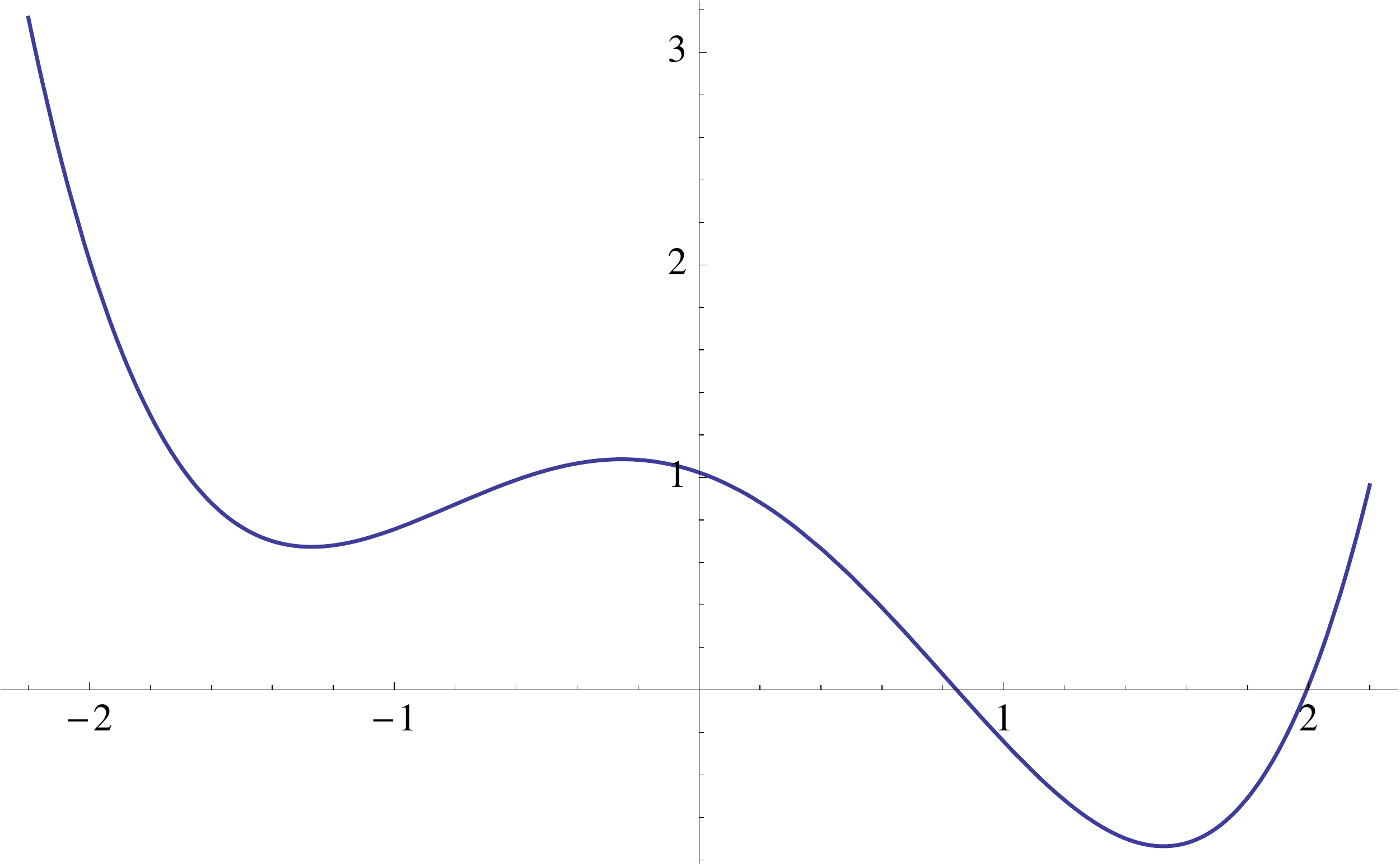}
\end{center}
\caption{Left panel: $g=g_{\rm crit}-0.5$ where $g_{\rm crit}=0.544331$; this
is a double well. Right panel: $g=g_{\rm crit}-0.3$ where $g_{\rm crit}=
0.544331$; the barrier between the wells is disappearing.}
\label{F56}
\end{figure}

\begin{figure}[h!]
\begin{center}
\includegraphics[scale=0.26]{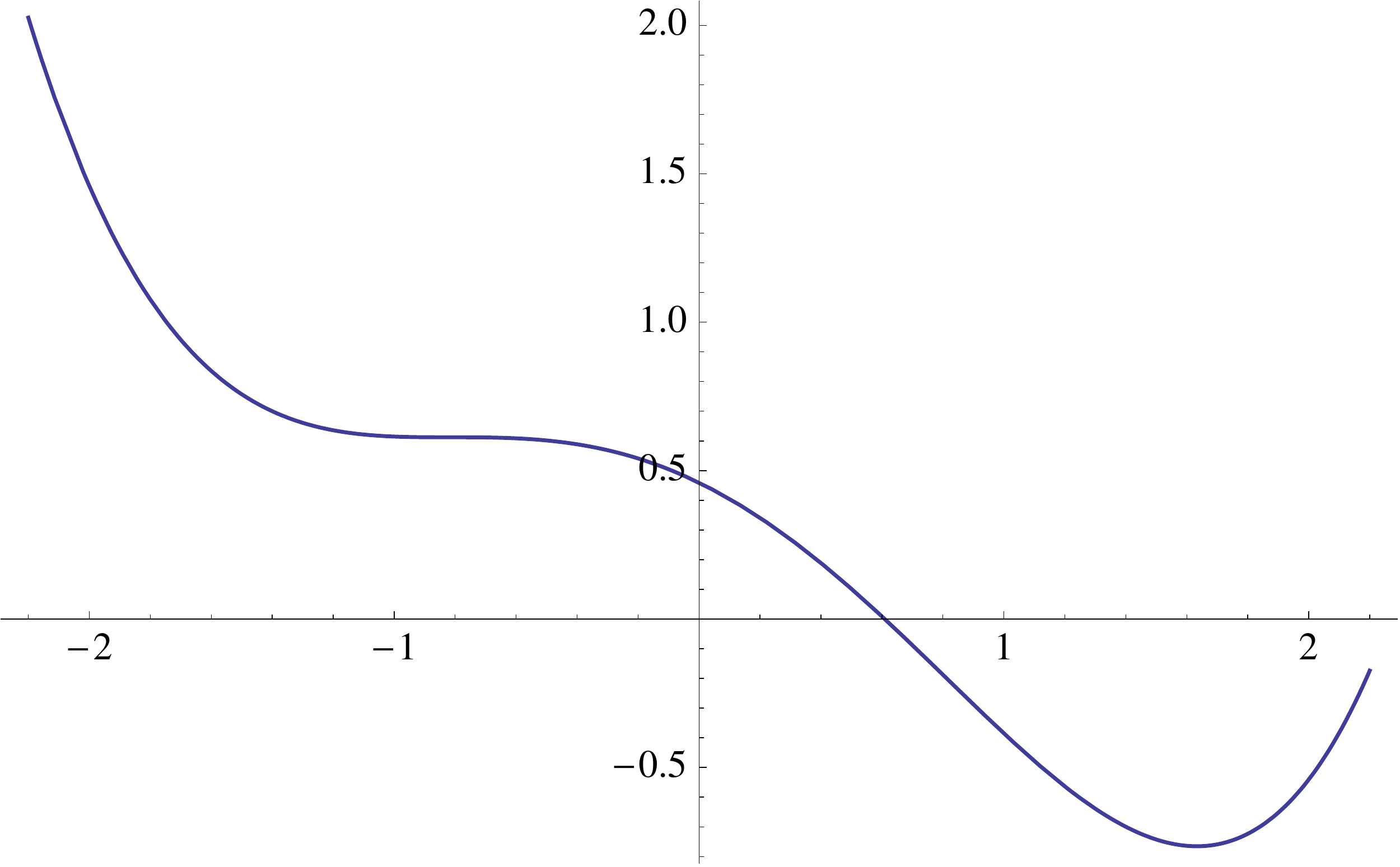}
\end{center}
\caption{$g=g_{\rm crit}$ where $g_{\rm crit}=0.544331$; the barrier between the
wells is gone and there is just a shoulder. The double well has become a single
well.}
\label{F7}
\end{figure}

Finally, we eliminate the one-derivative term and convert to an equation of
Schr\"odinger type. To do so, we substitute $\tilde\phi(s)=A(s)\chi(s)$,
which gives
$$\tilde\phi'(s)=A'(s)\chi(s)+A(s)\chi'(s)\quad{\rm and}\quad
\tilde\phi''(s)=A''(s)\chi(s)+2A'(s)\chi'(s)+A(s)\chi''(s).$$
Demanding that the $\chi'(s)$ terms drop out gives the following condition on
$A(s)$: 
\begin{equation}
A'(s)=\left(\frac{s^2}{8N^2g}-\frac{1}{4g}-1\right)A(s).
\label{e14}
\end{equation}
The equation for $\chi(s)$ then reduces to
\begin{equation}
-4g\chi''(s)+\left(\frac{1-2\beta}{2N^2}s+\frac{1}{16gN^4}s^4-\frac{1}{4gN^2}
s^2+\frac{1}{4g}\right)\chi(s)=\cE\chi(s).
\label{e15}
\end{equation}
Making the replacement $s\to sN$, we get
\begin{equation}
-\frac{4g}{N^2}\chi''(s)+\left[\frac{1-2\beta}{2N}s+\frac{1}{4g}\left(\frac{1}
{2}s^2-1\right)^2\right]\chi(s)=\cE\chi(s)
\label{e16}
\end{equation}
and for large $N$, $\beta=N/2$. The derivative operator on the left side of
this equation is the equivalent (isospectral) Hermitian Hamiltonian, which
has a {\it positive} quartic term, and we identify the effective potential in
the large-$N$ limit as
\begin{equation}
V(s)=-\frac{s}{2}+\frac{1}{4g}\left(\frac{1}{2}s^2-1\right)^2.
\label{e17}
\end{equation}
The condition for criticality \cite{R7} is $V'(s)=V''(s)=0$. This condition
reproduces the result in (\ref{e7}) with the appropriate change in sign; namely,
that $g_{\rm crit}=(2/3)^{3/2}\approx0.544331\,$. (See Figs.~\ref{F56} and
\ref{F7}.)

Let us now examine the effective potential near the critical point and determine
the universal function that arises in the double-scaling limit. To do so, we
study the behavior of the potential $V(s)$ in (\ref{e18}) near the critical
point $g=g_{\rm crit}$. We let $s=s_{\rm crit}+\Delta$ and $g=g_{\rm crit}+G$
with $G$ small and negative. For nonzero $G$ the potential has two local minima
at $s=s_1$ and $s=s_3$ and a local maximum at $s=s_2$. These extremal points
satisfy $s_1<s_2<s_3$. At the critical point $g=g_{\rm crit}$ the two extrema
at $s=s_1$ and $s=s_2$ coalesce.

It is convenient to write $G=-\delta^2$, where $\delta<<1$. Away from
criticality the positions of the extrema are determined by the condition
$\frac{\partial}{\partial s}V\left(s_{\rm crit}+\Delta,g_{\rm crit}-\delta^2
\right)=0$. Hence,
\begin{equation}
2\delta^2+\Delta^2\left(-\sqrt{6}+\Delta\right)=0.
\label{e18}
\end{equation}
We seek a perturbative expansion for $\Delta$ in the form $\Delta=\sum_{n=1}^{
\infty}a_n\delta^n$, where the coefficients $a_n$ are real. Substituting this
series for $\Delta$ into (\ref{e18}) allows us to calculate the coefficients
$a_1=\pm(2/3)^{1/4}$ and $a_2=1/6$. This gives
$$\Delta=\Delta_\pm=\pm\left(\frac{2}{3}\right)^{1/4}\delta+\frac{1}{6}
\delta^2+\ldots\,.$$
At the extrema $s_1=s_{\rm crit}+\Delta_-$ and $s_2=s_{\rm crit}+\Delta_+$ the
potential takes the values
$$V\left(s_{\rm crit}+\Delta_\pm\right)=\sqrt{\frac{3}{8}}+\frac{3}{8}
\delta^2\pm\frac{1}{2}\left(\frac{3}{2}\right)^{1/4}\delta^3+\ldots\,.$$

The mean value of the potential between $s_1$ and $s_2$ is approximately
$\sqrt{3/8}-3G/8$. We take $\cE=\sqrt{3/8}-3G/8$ and, since we are studying the
$N\to\infty$ limit, this value of $\cE$ represents a high-lying level. The
turning points at this energy are roots of
\begin{equation}
V\left(s,g_{\rm crit}+G\right)-\cE=0.
\label{e19}
\end{equation}
These roots have the form $s=s_{\rm crit}+b_1\delta+b_2\delta^2+b_3\delta^3
+\ldots$. Thus, (\ref{e19}) has the form
\begin{eqnarray}
&&V\left(s_{\rm crit}+b_1\delta+b_2\delta^2+b_3\delta^3,g_{\rm crit}-\delta^2
\right)-\cE\nonumber\\
&&\qquad=\frac{3}{8}\left(\sqrt{6}b_1-b_1^3\right)\delta^3+\frac{3}{64}
\left(6^{3/2}+\sqrt{6}b_1^4+8\sqrt{6}b_2-24b_1^2b_2\right)\delta^4+\ldots\,.
\nonumber
\end{eqnarray}
From this equation we determine that the three possible values for $b_1$ are
\begin{equation}
b_1=6^{1/4},~0,~{\rm and}~-6^{1/4}.
\label{e20}
\end{equation}
Since the formula for $b_2$ in terms of $b_1$ is
$$b_2=-\frac{\sqrt{6}}{8}\frac{\left(6+b_{1}^4\right)}{\sqrt{6}-3b_1^2},$$
the values of $b_2$ corresponding to those for $b_1$ from (\ref{e20}) are
\begin{equation}
b_2=3/4,~-3/4,~{\rm and}~3/4.
\label{e21}
\end{equation}
The resulting values for the turning points in $s$ are $s_-<s_0<s_+$, where
$$s_0=s_{\rm crit}+\frac{3}{4}G+\ldots\quad{\rm and}\quad s_\pm=s_{\rm crit}-
\frac{3}{4}G\pm 6^{1/4}\left(-G\right)^{1/2}+\ldots\,.$$

Near the critical point $V\left(s,g_{\rm crit}+G\right)-\cE$ can be approximated
by $-\frac{3}{8}s\left(s^2+6^{1/2}G\right)$ to leading order in $\left(-G\right)
^{1/2}$, where we have made a shift in the origin of $s$ to $s_{\rm crit}$.
Hence, in the neighborhood of the critical point the Schr\"odinger equation
reads
\begin{equation}
\frac{4g}{N^2}\frac{d^2\chi}{ds^2}+\frac{3}{8}s\left(s^2+6^{1/2}G\right)\chi=0.
\label{e22}
\end{equation}
On making the change of variable $t=6^{1/4}(-G)^{-1/2}s$, we obtain the
Schr\"odinger equation valid near the critical point $g=g_{\rm crit}$:
\begin{equation}
-\chi''(s)+\gamma s\left(1-s^2\right)\chi(s)=0,
\label{e23}
\end{equation}
where
\begin{equation}
\gamma=\frac{27}{64}6^{3/4}N^2(-G)^{5/2}.
\label{e24}
\end{equation}
Thus, the double-scaling limit is obtained by letting $G\to0$ and $N\to\infty$
in such a way that $N\left(-G\right)^{5/4}$ is held fixed. Because the universal
function $\chi(s)$ in this double-scaling limit satisfies the differential
equation (\ref{e24}), which has no singular points in the finite complex-$s$
plane, $\chi(s)$ is an entire function of $s$.

\section{Discussion and conclusions}
\label{s4}

In this paper we have identified and clarified a problem with the double-scaling
limit in a prototypical quantum field theory in one-dimensional spacetime;
namely, that the critical coupling constant $g_{\rm crit}$ in the double-scaling
limit is {\it negative}. We have argued that for a conventional Hermitian
one-dimensional quartic theory the double-scaling limit does not exist because
near $g_{\rm crit}$ the potential is upside-down and thus the wave function is
not normalizable; particles are not confined and can tunnel out to $r=\infty$.
However, we have shown that if we approach the critical theory in a $\cP
\cT$-symmetric fashion, the resulting double-scaling limit gives a physically
acceptable quantum theory and this theory is equivalent (isospectral) to a
conventional Hermitian quantum theory with a confining potential.

We are not interested in the complex quantum theory that one obtains by
analytically continuing the coupling constant $g$ in the potential $V=x^2+gx^4$
from positive $g$ to negative $g$ in the complex-$g$ plane. (The eigenvalues
of the resulting Hamiltonian are complex. Worse yet, they are not unique because
they are path dependent; one obtains different eigenvalues depending on whether
one rotates around $g=0$ in a clockwise or an anticlockwise direction
\cite{R4}.) Rather, our approach is to keep the coupling constant $g$ fixed and
to perform the $\cP\cT$-symmetric limit of $x^2+gx^2(ix)^\epsilon$ as $\epsilon$
goes from $0$ to $2$. The integration path lies on the real axis when $\epsilon=
0$ and the path rotates downward into the complex plane as $\epsilon$ increases.
When $\epsilon=2$, the contour comes inward from $x=\infty$ in the $60^\circ$
Stokes wedge $-\pi<{\rm arg}\,x<-2\pi/3$ and goes back out to $x=\infty$ in the
$60^\circ$ Stokes wedge $-\pi/3<{\rm arg}\,x<0$. As a consequence, the
eigenfunctions are normalizable. In this paper we have used O($N$) symmetry to
perform the $N$-dimensional version of this $\cP\cT$-symmetric analytic
continuation, thus to obtain the quartic theory that is studied in this paper.

It is interesting that in zero-dimensional spacetime we found in Ref.~\cite{R3}
that the double-scaling limit is characterized by the asymptotic behavior $G\sim
N^{-1/3}$, while in one-dimensional spacetime we have shown in this paper that
$G\sim N^{-4/5}$. We do not know yet what happens in higher dimensional
spacetime. Stokes wedges represent {\it global} constraints and understanding
the structure of these wedges becomes very difficult when the dimension of
spacetime increases \cite{R5}. Many studies of the double-scaling limit in field
theory rely on applying formal saddle-point methods to functional integrals and
these techniques involve only {\it local} analysis.

$\cP\cT$-symmetric field theories, at least in low dimension, provide an arena
in which the double-scaling limit can be performed consistently. We have used a
different methodology in the one-dimensional spacetime case from the
zero-dimensional spacetime case. To extend our work to $D$-dimensional spacetime
with $D\geq2$, we will need to develop even more powerful methods. The usual
heuristic treatment of functional integrals is inadequate to discuss the global
questions addressed in this paper. In future work we shall pursue two different
research directions: (i) $\cP\cT$-symmetric versions of matrix models
\cite{R10}, and (ii) Schwinger-Dyson equations for $\cP\cT$-symmetric field
theories for $D\geq2$ \cite{R11}.

\vspace{0.5cm}
\footnotesize
\noindent
We thank the U.K. Royal Society for grant support under the International
Exchanges Scheme. CMB was supported in part by the U.S.~Department of Energy.

\end{document}